\documentclass[twocolumn,aps,pra,showpacs]{revtex4}

\begin{document}
\title{Solving the liar detection problem using the four-qubit singlet state}
\author{Ad\'{a}n Cabello}
\email{adan@us.es}
\affiliation{Departamento de F\'{\i}sica Aplicada II,
Universidad de Sevilla, 41012 Sevilla, Spain}
\date{\today}


\begin{abstract}
A method for solving the Byzantine agreement problem [M. Fitzi, N.
Gisin, and U. Maurer, Phys. Rev. Lett. {\bf 87}, 217901 (2001)]
and the liar detection problem [A. Cabello, Phys. Rev. Lett. {\bf
89}, 100402 (2002)] is introduced. The main advantages of this
protocol are that it is simpler and is based on a four-qubit
singlet state already prepared in the laboratory.
\end{abstract}


\pacs{03.67.Hk,
02.50.Le,
03.65.Ud,
03.65.Ta}
\maketitle




\section{Introduction}
\label{sec:I}


Some interesting applications of the $N$-particle $N$-level
singlet states have been recently introduced \cite{Cabello02b}.
Particularly, the three three-level singlet state
\begin{eqnarray}
\left|{\cal S}_3^{(3)} \right\rangle & = & {1 \over \sqrt {6}} (
\left| 012 \right\rangle - \left| 021 \right\rangle -
\left| 102 \right\rangle + \left| 120 \right\rangle \nonumber \\
& & + \left| 201 \right\rangle - \left| 210 \right\rangle )
\label{s3s3}
\end{eqnarray}
(the lower and upper indexes refer, respectively, to the number of
constituents and the dimension of the Hilbert space of any of the
constituents of the composite system) was shown to be useful for
solving the ``liar detection problem'' \cite{Cabello02b}, which is
a simplification of a classically unsolvable problem in quantum
computing called the ``Byzantine generals' problem'' or the
``Byzantine agreement problem'' introduced in two seminal papers
by Lamport, Pease, and Shostak \cite{PSL80,LSP82}, and inspired by
a pseudohistorical scenario \cite{commh}. A version of this
problem was previously solved by Fitzi, Gisin, and Maurer
\cite{FGM01} also using the three three-level singlet state.
Ref.~\cite{Cabello02b} ended by remarking that the preparation of
the required states would pose a formidable challenge. Meanwhile,
Gisin has proposed a method for preparing the three three-level
singlet state \cite{Gisin02}. To my knowledge, no experimental
results have as yet been reported.

On the other hand, the Munich group \cite{Weinfurter02} (see also
Ref.~\cite{WZ01}) has prepared, using parametric down-converted
photons entangled in polarization, a four-qubit entangled state
that can be expressed as
\begin{eqnarray}
\left|{\cal S}_4^{(2)} \right\rangle & = & {1 \over 2 \sqrt{3}}
(2|0011\rangle-|0101\rangle-|0110\rangle \nonumber \\
& & -|1001\rangle-|1010\rangle+2|1100\rangle). \label{s4s2}
\end{eqnarray}
This state belongs to a family of states of $N$ (even) qubits that
generalizes the familiar two-qubit singlet state
\begin{equation}
\left|{\cal S}_2^{(2)} \right\rangle = {1 \over \sqrt{2}}
(|01\rangle-|10\rangle). \label{s2s2}
\end{equation}
Any member of this family can be expressed as
\begin{eqnarray}
\left|{\cal S}_N^{(2)} \right\rangle = {1 \over {N \over 2}! \sqrt
{{N \over 2}+1} }\sum_{\scriptscriptstyle{
{\stackrel{\scriptscriptstyle{\rm permutations}} {{\rm
of}\;0\ldots01\ldots1}}}} \!\!\!\!\!\! z!
\left( {N \over 2}-z \right)! (-1)^{{N \over 2}-z} \nonumber \\
\times \left| ij\ldots n \right\rangle, \label{SnS2}
\end{eqnarray}
where the sum is extended to all of the states obtained by
permuting the state $|0\ldots 01\ldots1\rangle$, which contains
the same number of 0's and 1's; $z$ is the number of 0's in the
first $N / 2$ positions (for example, in $|01\rangle$, $z=1$; in
$|1100\rangle$, $z=0$; in $|010110\rangle$, $z=2$). Expression
(\ref{SnS2}) is similar to that introduced in
Ref.~\cite{Cabello02b} for the $N$-particle $N$-level singlet
states (which, according to the notation introduced here, should
be denoted as $|{\cal S}_N^{(N)} \rangle$).

The interesting point is that the $N$-qubit ``singlet''
\cite{comm1} states given by Eq.~(\ref{SnS2}) share some
properties with the $N$-particle $N$-level singlet states
introduced in Ref.~\cite{Cabello02b}. Both are nonseparable and
$N$-lateral unitary invariant. Nonseparability means that no local
model can mimic the predictions of quantum mechanics for these
states \cite{comm2} and implies that the outcomes of the
measurements do not reveal predefined results. $N$-lateral unitary
invariance means that, if we act on any state with the tensor
product of $N$ equal unitary operators, the result will be to
reproduce the same state:
\begin{equation}
U^{\bigotimes N} \left|\psi\right\rangle= \left|\psi\right\rangle,
\end{equation}
$U^{\bigotimes N}$ being $U \otimes \ldots \otimes U$, where $U$
is a unitary operator. This implies that the same correlations
between the outcomes of measurements on the $N$ particles occur
for different sets of measurements. Both properties were essential
to the solutions to the problems proposed in
Refs.~\cite{Cabello02b,FGM01}. The question is whether the $|{\cal
S}_N^{(2)}\rangle$ states can be used to perform tasks that, so
far, were specific to the~$|{\cal S}_N^{(N)}\rangle$ states.

The $|{\cal S}_N^{(2)}\rangle$ states can be used to distribute
cryptographic keys \cite{BB84,Ekert91}, encode quantum information
in decoherence-free subspaces \cite{PSE96,DG97,ZR97a,KBLW01},
perform secret sharing, and teleclone quantum states (telecloning
is a process combining quantum teleportation and optimal quantum
cloning from $1$ input to $M$ outputs) \cite{MJPV99}. In fact, for
$N \ge 4$, the $|{\cal S}_N^{(2)}\rangle$ states are the
$N$-lateral unitary invariant version of the telecloning states
introduced in Ref.~\cite{MJPV99}. In this paper, I shall show that
the state $|{\cal S}_4^{(2)} \rangle$ can also be used to solve
the Byzantine generals' and liar detection problems.

The {\em Byzantine generals' problem} is a classical problem in
distributed computing defined as follows. $N$~generals are
connected by secure pairwise classical channels. A commanding
general must send an order to his $N-1$~lieutenant generals such
that: (a) All loyal lieutenant generals obey the same order; (b)
if the commanding general is loyal, then every loyal lieutenant
general obeys his order. If the commanding general is loyal, (a)
follows from (b). However, the commanding general may be a
traitor.

The Byzantine generals' problem is a metaphor for distributed
processes, some of which may be faulty. For $N=3$~generals and
$1$~traitor the problem has no classical solution, as proved in
Refs.~\cite{PSL80,LSP82}. However, a version thereof can be solved
with the aid of quantum mechanics. The protocol proposed in
Ref.~\cite{FGM01} solves the Byzantine generals' problem for $N=3$
in the following sense: if all generals are loyal then, after the
protocol, all the generals would obey the commanding general's
order. If one general is a traitor then, after the protocol,
either the two loyal generals would obey the commanding general's
order or abort the process.

As can be easily seen, the Byzantine generals' problem for
$N=3$~generals and $1$~traitor is equivalent to the {\em liar
detection problem} defined as follows \cite{Cabello02b}: three
parties $A$(lice), $B$(ob), and $C$(arol) are connected by secure
pairwise classical channels. $A$ sends a message $m$ to $B$ and
$C$, and $B$ forwards the message to $C$. If both $A$ and $B$ are
honest, then $C$ should receive the same message from $A$ and $B$.
However, $A$ could be dishonest and send different messages to $B$
and $C$, $m_{AB} \neq m_{AC}$, or, alternatively, $B$ could be
dishonest and send a message that is different from that he has
received, $m_{BC} \neq m_{AB}$. For $C$ the problem is to
ascertain without a shadow of a doubt who is being dishonest.

In Sec.~\ref{sec:III} I shall introduce a protocol based on
the~$|{\cal S}_4^{(2)} \rangle$ state for solving the liar
detection problem (and thus also the Byzantine generals' problem)
in the following sense: if both $A$ and $B$ are honest then $C$
will receive the same message from $A$ and $B$. If one of $A$ and
$B$ is dishonest then, after the protocol, either $C$ ascertain
who is being dishonest or $C$ does not accept the message from one
of the parties $A$ and $B$.

Before getting into the details of the protocol, it would be
useful to give a rough idea of how it works: besides their
messages, $A$ and $B$ must send some additional information. For
instance, $A$ must also send $B$ some private information
correlated with the message. $B$ and $C$ must be able to check the
authenticity of the information they have received by using
private information. In addition, to convince $C$ that the message
$B$ is sending her is actually that $B$ received from $A$, $B$
would also need to send~$C$ the information $B$ has received from
$A$. In this scenario, being a liar means that she/he did not
distribute the appropriate additional information. The advantage
derives from the fact that $C$ can detect the origin of the
inappropriate information.


\section{Distribute and test protocol}
\label{sec:II}


The protocol for solving the liar detection problem described in
Sec.~\ref{sec:III} is based on the assumption that, at the
beginning of the protocol, $A$, $B$, and $C$ share a large number
$L$ of four-qubit systems in the $|{\cal S}_4^{(2)} \rangle$ state
described by Eq.~(\ref{s4s2}) such that, for each four-qubit
system $j$ (from 1 to $L$), $A$ possesses two qubits (qubits $j_1$
and $j_2$, or qubits $j_1$ and $j_3$, but she does not know which
two), $B$ possesses one qubit ($j_3$ or $j_2$, but he does not
know which one), and~$C$ possesses the fourth qubit $j_4$ and
knows which qubits $A$ and $B$ have in their possession.

The purpose of this distribute and test protocol, preceding the
protocol for liar detection, is to prepare the initial scenario
required for the second protocol and test whether or not said
initial scenario has been achieved. The distribute and test
protocol has only two possible outcomes: success or failure. In
case of success, $C$ would assume that the required initial
scenario has been achieved and then the liar detection can be
reliably accomplished. In case of failure, $C$ would conclude that
something went wrong and would abort any subsequent action.

The initial assumptions are that there are only three parties
involved: $A$, $B$, and $C$, and that they are connected by secure
pairwise classical channels and by noiseless quantum channels.

The distribute and test protocol is as follows.

(i) $C$ prepares a large number $M>L$ of four-qubit systems in the
$|{\cal S}_4^{(2)} \rangle$ state. For each four-qubit system $j$
(from 1 to $M$), $C$ sends two qubits (qubits $j_1$ and $j_2$, or
qubits $j_1$ and $j_3$) to $A$, one qubit to $B$ ($j_3$ or $j_2$),
and keeps the fourth qubit for herself.

(ii) For each four-qubit system $j$ sent, $C$ checks whether or
not $A$ and $B$ have received the right number of qubits (two
qubits for $A$ and one for $B$).

(iii) $C$ randomly chooses two large enough subsets $S_1$ and
$S_2$ of the distributed four-qubit systems. These subsets are
used only to test whether or not the initial scenario required for
the protocol for liar detection has been achieved. Qubits
belonging to $S_1$ and $S_2$ will be discarded at the end of the
distribute and test protocol. If $N_i$ is the number of four-qubit
systems in $S_i$, then $M=N_1+N_2+L$.

(iv) For each four-qubit system $j$ of $S_1$, $C$ asks $A$ to send
$B$ her two qubits.

(v) $C$ checks whether or not $B$ has received $A$'s qubits.

(vi) $C$ asks $B$ to perform some measurements on each of the
three qubits (the two qubits $B$ has received from $A$ plus the
one that has been in $B$'s possession from the start), and report
his results to her by using the secure pairwise classical channel
between them. For instance, $C$ can ask $B$ to perform a
measurement of the spin along the same direction on all three
qubits, $M_j$.

(vii) $C$ performs a measurement on her corresponding qubit. For
instance, $C$ can measure $M_j$. The outcomes of $B$ and $C$'s
measurements must be correlated. For instance, if they have
measured $M_j$ on the four qubits, two of them must be ``$0$'' and
the other two must be ``$1$''; due to unitary invariance, this
occurs for any $M_j$. Due to the fact that $A$'s qubits are
entangled with $B$ and $C$'s qubits, the outcomes can be proved
\cite{comm2} to be genuinely unpredictable (if one does not know
the results of the other measurements). $C$ checks whether the
expected correlations occur.

(viii) For each four-qubit system $j$ of $S_2$, $C$ follows the
same steps as in (iv)--(vii), only exchanging the roles of $A$ and
$B$.

(ix) Since $C$'s choice of subsets $S_1$ and $S_2$ is known by $A$
and $B$ only after all qubits have been distributed among them,
then, if all the outcomes are correctly correlated, $C$ would
assume that the remaining $L$ distributed four-qubit systems are
in the $|{\cal S}_4^{(2)} \rangle$ state and use them for the
protocol described below. If not, $C$ would conclude that
something went wrong and abort any subsequent action.


\section{Protocol for liar detection}
\label{sec:III}


Let us suppose that the message $m$ that $A$ sends to $B$ and $C$,
and $B$ sends to $C$ is a bit value ``$0$'' or ``$1$'' and that
all three parties agree to use the following protocol.

(I) For each four-qubit system, $C$ asks $A$ and $B$ to perform
the same measurement on their individual qubits. After a large
number of these measurements, both $B$ and $C$ ($A$) are in
possession of a long list of (pairs of) 0's and 1's $l_B$, and
$l_C$ ($l_A$) such as the following.

\begin{table}[tbph]
\begin{center}
\begin{tabular}{lccc}
\hline \hline \multicolumn{1}{c}{Position} & $l_A$ & $l_B$ & $l_C$ \\
\hline
$1$ & $00$ & $1$ & $1$ \\
$2$ & $01$ & $0$ & $1$ \\
$3$ & $00$ & $1$ & $1$ \\
$4$ & $11$ & $0$ & $0$ \\
$5$ & $11$ & $0$ & $0$ \\
$6$ & $00$ & $1$ & $1$ \\
$7$ & $01$ & $0$ & $1$ \\
$8$ & $11$ & $0$ & $0$ \\
$\ldots$ & $\ldots$ & $\ldots$ & $\ldots$ \\
\hline \hline
\end{tabular}
\end{center}
\end{table}

Each of these lists has the following properties.

(a) It is random (i.e., generated by an intrinsically unrepeatable
method that gives each possible number the same probability of
occurring).

(b) It is correlated to the other two lists. If ``$00$''
(``$11$'') is in position $j$ in $l_A$, then ``$1$'' (``$0$'') is
in position $j$ in $l_B$ and $l_C$.

(c) Each party knows only his/her own list.


(II) The message $A$ sends $B$ is denoted as $m_{AB}$. In addition
to $m_{AB}$, $A$ must also send $B$ the list $l_{A}^{(m_{AB})}$ of
positions in $l_A$ in which $m_{AB}$ appears twice. For instance,
if $A$ sends $B$ the message $m_{AB}=0$, then she must also send
$B$ the list $l_{A}^{(m_{AB}=0)}=\{1,3,6,\ldots\}$ since, in
$l_A$, ``$00$'' appears in positions 1, 3, 6,\ldots Note that, if
the sequences are random and long enough, then the length of
$l_{A}^{(m_{AB}=0)}$ must be about one-quarter of the total length
$L$ of the list $l_A$.

(III) $B$ would not accept the message if the received list
$l_{A}^{(m_{AB})}$ is not compatible with $l_B$ or the length of $
l_{A}^{(m_{AB})} \ll L/4$. For instance, if
$l_{A}^{(m_{AB}=0)}=\{1,2,3,6,\ldots\}$, then $B$ would not accept
the message because ``$0$'' is in position 2 in $l_B$, so $A$
cannot have ``$00$'' in this position.

Requirements (II) and (III) force $A$ to send $B$ information
which is correct but perhaps incomplete. Otherwise, if $A$ sends a
list containing $n$ erroneous data, then the probability that $B$
does not accept the message $m_{AB}$ would be at last
$(2^n-1)/2^n$.

(IV) The message $B$ sends $C$ is denoted as $m_{BC}$. $B$ must
also send $C$ a list $l_{A}^{(m_{BC})}$, which is supposedly the
same $l_{A}^{(m_{AB})}$ that $B$ has received from $A$.

(V) The message $A$ sends $C$ is denoted as $m_{AC}$. $A$ must
also send $C$ a list $l_{AC}$, which is supposedly $l_A$.

(VI) When $C$ finds that $m_{AC} \neq m_{BC}$, she has three lists
$l_C$, $l_{A}^{(m_{BC})}$, and $l_{AC}$, to help her find out
whether it is $A$ or $B$ who is being dishonest. $C$ must first
check whether $l_{AC}$ is consistent with $l_C$. If not, then $A$
is the liar. If yes, then $C$ must check whether
$l_{A}^{(m_{BC})}$ has an appropriate length and is consistent
with $l_{AC}$. If not, then $B$ is the liar. At this stage, this
is the only possibility.


\section{Conclusions}
\label{sec:IV}


The main practical advantage of the introduced protocol over those
presented in Refs.~\cite{Cabello02b,FGM01} is that it requires a
simpler quantum state (a four-qubit state instead of a
three-qutrit state) which has been already prepared in a
laboratory \cite{Weinfurter02}. This would make the new protocol
immediately applicable for solving both the Byzantine generals'
and liar detection problems. The main theoretical advantage is
that the introduced protocol seems to be fundamental in a greater
degree than those in Refs.~\cite{Cabello02b,FGM01}, in the sense
that it requires that only one party uses trit values, instead of
all three parties.

Although both the Byzantine generals' and liar detection problems
assume that the parties are connected by secure pairwise classical
channels (secure messengers in Ref.~\cite{commh}), this is an
unrealistic constraint for real applications. Note, however, that
the protocol described above also works if the classical channels
between $A$ and $C$, and between $B$ and $C$ are not secure but
public and unjamable (i.e., which cannot be tampered with).
Obviously, such a channel cannot be used to distribute delicate
information (like, for instance, whether or not the generals will
attack, or the lists $l_{A}^{(m_{BC})}$ and $l_{AC}$), but can be
used, together with an additional quantum channel, to implement a
standard quantum key distribution protocol \cite{BB84,Ekert91} to
send this delicate information. The final picture gives us a
quantum solution to a problem without classical solution: the liar
detection problem without secure classical channels.


\begin{acknowledgments}
I thank N.~Gisin for useful suggestions, M.~Bourennane, M.~Eibl,
S.~Gaertner, N.~Kiesel, C.~Kurtsiefer, and H.~Weinfurter for
discussions, C.~Serra and M.~\.{Z}ukowski for comments,
H.~Weinfurter for his hospitality at
Ludwig-Maximilians-Universit\"{a}t, M\"{u}nchen and
Max-Planck-Institut f\"{u}r Quantenoptik, Garching, and the
Spanish Ministerio de Ciencia y Tecnolog\'{\i}a Grants
Nos.~BFM2001-3943 and BFM2002-02815, the Junta de Andaluc\'{\i}a
Grant No.~FQM-239, the Max-Planck-Institut f\"{u}r Quantenoptik,
and the European Science Foundation (Quantum information theory
and quantum computation short scientific visit grant) for support.
\end{acknowledgments}



\begin{thebibliography}{99}
\bibitem{comm1} The name ``singlet'' for the
$|S_2^{(2)}\rangle$ state, the only two-qubit state of total spin
zero, was coined, together with the name ``triplet'' (which
denotes three orthogonal states of total spin one). In this paper,
I use the name ``singlet'' for any of the $|S_N^{(2)}\rangle$
states because, as Eq.~(\ref{SnS2}) shows, they are a
generalization of $|S_2^{(2)}\rangle$. However, for $N>2$, with
$N$ even, the dimension of the subspace of total spin zero is $N!
/ [(N/2)! (N/2+1)!]$ (see Ref.~\cite{ZR97a}).
\end{thebibliography}
\end{document}